# Three-Sensor 2ω Method with Multi-directional Layout: A General Methodology for Measuring Thermal Conductivity of Solid Materials

Guang Yang, Bing-yang Cao*

*Key Laboratory for Thermal Science and Power Engineering of Ministry of Education, Department of Engineering Mechanics, Tsinghua University, Beijing 100084, People's Republic of China*

**Abtract:**

Anisotropic thermal transport plays a key role in both theoretical study and engineering practice of heat transfer, but accurately measuring anisotropic thermal conductivity remains a significant challenge. To address this issue, we propose the three-sensor 2ω method in this study, which is capable of accurately measuring the isotropic or anisotropic thermal conductivity of solid materials. In this method, several three-sensor groups following the design guidelines are fabricated upon the sample along different characteristic directions, and each group consists of three parallel metal sensors with unequal widths and distances optimally designed based on sensitivity analysis. Among the three sensors, the outer two serve as AC heaters and the middle one as a DC detector. The 2ω voltage signals across the detector in each three-sensor group are measured, and then the data are processed by the proposed Intersection Method to derive the thermal conductivities along directions of interest. The application of the detector's 2ω instead of the heater's 3ω voltage signals eliminates the errors introduced by the uncertainties of thermal resistance in superficial structures (metal layer, insulation layer, interface, etc.). Meanwhile, by replacing the fitting algorithm with the Intersection Method, the local optimum trap of multivariate fitting is avoided. To verify the accuracy and reliability, four typical monocrystalline semiconductors, i.e., Si, GaN, AlN, and $\beta$-Ga$_2$O$_3$, are measured, and the results are consistent with the literature. This method will provide a comprehensive and versatile solution for the thermal conductivity measurements of solid materials.

* Corresponding author: E-mail: caoby@mail.tsinghua.edu.cn

# 1. Introduction

Anisotropic heat transport phenomena are ubiquitous in materials, and are crucial to the theoretical study and technological development of heat transfer[1]. On the one hand, if the lattice structure of a material is highly asymmetric, it generally exhibits a significant anisotropic thermal conductivity. On the other hand, if a material with a symmetric lattice contains highly oriented lattice defects (e.g., dislocations, grain boundaries, stacking faults), it tends to exhibit an anisotropic thermal conductivity likewise[2, 3]. For various advanced technology fields, the anisotropic thermal conductivity is one of the key indicators that determine the performance and reliability, especially in the thermal management of electronic devices[4], thermal rectification[5], high-temperature superconductor[6], thermal photovoltaics[7, 8], and thermoelectricity[9, 10], etc. In recent years, researchers have made considerable progress in developing experimental techniques for thermal conductivity measurement, however, accurately measuring the anisotropic thermal conductivity of solid materials remains a challenge.

The experimental techniques available for measuring the anisotropic thermal conductivity of solid materials include two categories: optical and electrical methods. The optical methods mainly include the ultrafast laser transient thermoreflectance (TTR, including the original time-domain thermoreflectance, TDTR[11-14], and the original frequency-domain thermoreflectance, FDTR[15, 16]), and various improved laser pump-probe methods. Among the improved pump-probe methods, representative works involve the asymmetric-beam time-domain thermoreflectance (AB-TDTR[3, 17], or the elliptical-beam TDTR[18, 19]), the beam-offset time-domain thermoreflectance (BO-TDTR)[20, 21], the beam-offset frequency-domain thermoreflectance (BO-FDTR)[2, 22-25], and the spatial-domain thermoreflectance (SDTR)[26], etc. The most remarkable advantage of the optical methods is the flexibility to adjust the sample's orientation angle (or the relative positions of pump and probe laser spots), and the elliptical eccentricity of the laser spot, thereby directly achieving a 360° scan of the in-plane anisotropic thermal conductivity[2, 19].

Nevertheless, there are two inevitable limitations of these optical methods. First, probe signals of these optical methods (except the SDTR method) are generally sensitive to the thermophysical properties of superficial structures, i.e., thermal conductivity and heat capacity of the metal transducer, and transducer-sample thermal boundary resistance (TBR), owing to the laser's high repeating frequency, and this introduces additional errors in the thermal conductivity measurements. Meanwhile, these methods depend on multivariate fitting algorithms to derive the undetermined thermophysical properties, which are difficult to guarantee that the fitting results converge to the global optimum, introducing incalculable errors in the final measurement results. Regarding the novel SDTR method adopting low frequency, it is problematic to measure the cross-plane thermal conductivity due to insufficient sensitivity[26], despite SDTR overcoming the first of the two limitations of other optical methods.

In terms of electrical methods, the main ones applied to measure anisotropic thermal conductivity are the various harmonic methods developed from the classical 3ω method[27, 28], including the multi-sensor 3ω method[29-36], the suspended 3ω method[28, 29, 37], and the two-sensor 2ω method[38-40]. Microfabricated devices[41, 42] are also commonly used for anisotropic thermal conductivity measurements of low-dimensional materials. However, all the above electrical methods (except the two-sensor 2ω method) are affected by the error propagation from the thermal resistance of the superficial structures (insulation layer, sensor-sample interface, etc.) to the final measurement results[38, 39]. And similar to the optical methods, these electrical methods (except the microfabricated devices) also depend on multivariate fitting algorithms, which seriously challenges the accuracy of the measured results. Moreover, it is difficult for the suspended 3ω method and the microfabricated devices to measure the cross-plane thermal conductivity, unless one resorts to the classical 3ω method or the differential 3ω method[37, 41].

Table 1 summarizes the comparison of currently available experimental methods. It is evident that a high-precision experimental method with universal applicability for measuring the isotropic or anisotropic thermal conductivities of solid materials is still lacking. Fortunately, it has been found in our prior work[43] and other related works[38, 39] that the second harmonic (2ω) voltage signals across a DC detector is insensitive to the thermophysical properties of superficial structures, with the detector located near an AC heater. In this work, we propose the three-sensor 2ω method, which replaces multivariate fitting algorithms by the proposed Intersection Method, and inherits the spirit of 2ω signals to eliminate the error propagation due to uncertainties in the thermal resistance of samples' superficial structures. Thus, the proposed three-sensor 2ω method effectively overcomes the limitations of the existing methods listed in Table 1.

| Category | Method | Sensitive to both cross- and in-plane thermal conductivities? | No multivariate fitting? | Invulnerable to the errors introduced by superficial structures? | Refs. |
|---|---|---|---|---|---|
| Optical | Original TDTR/FDTR | Yes | No | No | [11-16] |
| | AB-TDTR | Yes | No | No | [3, 17-19] |
| | BO-TDTR/BO-FDTR | Yes | No | No | [2, 20-25] |
| | SDTR | No | No | Yes | [26] |
| Electrical | Multi-sensor 3ω | Yes | No | No | [29-36] |
| | Suspended 3ω | No | No | No | [28, 29, 37] |
| | Two-sensor 2ω | Yes | No | Yes | [38-40] |
| | Microfabricated devices | No | Yes | No | [41, 42] |
| | **Three-sensor 2ω** | **Yes** | **Yes** | **Yes** | **This work** |

**Table 1.** Comparison of currently available techniques for measuring anisotropic thermal conductivity.

This paper is organized as follows. First, the experimental system and measurement procedure (viz, the Intersection Method) are illustrated. Then the required guidelines for the three-sensor layout design are highlighted based on the signal sensitivity analysis. Finally, measurements are conducted on four monocrystalline semiconductor samples (i.e., Si, GaN, AlN, and $\beta$-Ga$_2$O$_3$) to verify the accuracy and applicability of the proposed three-sensor 2ω method.

## 2. Method
### 2.1 Experimental Setup

Figure 1 demonstrates the experimental system and sample structure of the proposed three-sensor 2ω method. The method is applied to measure the thermal conductivity of solid materials, regardless of their isotropic or anisotropic nature. As shown in Figure 1(a), in order to prevent leakage and signal crosstalk between sensors, an insulation layer is deposited on the sample surface, except for well-insulating samples. Multiple three-sensor groups are then prepared on the surface along different characteristic directions of interest using lithography, sputtering, and lift-off processes. Each three-sensor group consists of three parallel metal sensors with different widths and distances (Figure 1(b)) optimally designed based on sensitivity analysis. The outer two sensors function as heaters, with the wider one referred to as heater 1 and the narrower one as heater 2. A detector is situated between the two heaters, consistent with our prior works[43, 44]. The substrate, insulation layer, and multiple three-sensor groups on the surface comprise the effective test sample, which allows for deriving thermal conductivities along different orientations of interest.

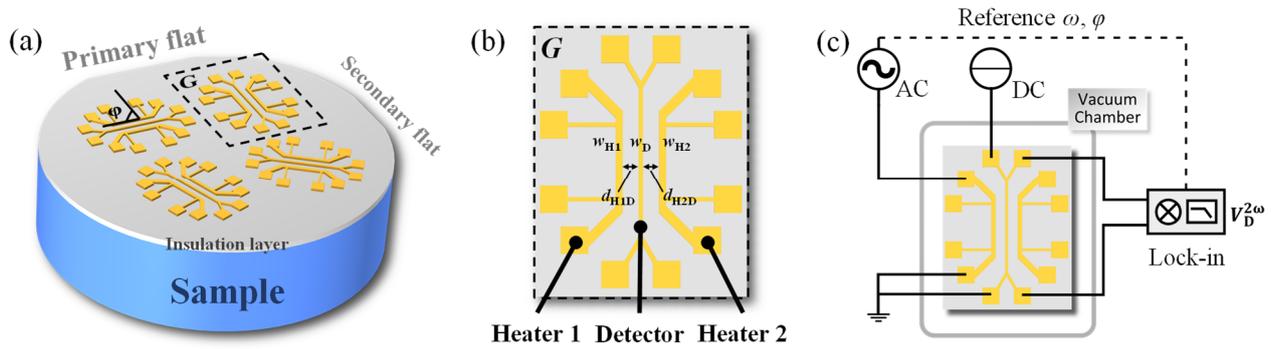

**Figure 1. Typical structure of a test sample and the experimental system of the three-sensor 2ω method.** (a) Structure of a test sample and layout of the three-sensor groups. $\angle\varphi$ indicates the angle between the length direction of a three-sensor group and a given direction (e.g., the wafer secondary flat in this figure). For example, the dashed region G indicates the three-sensor group whose length direction is perpendicular to the primary flat (i.e., $\angle\varphi = 0$). Several three-sensor groups can be arbitrarily prepared upon the insulation layer, with their length directions along the orientations of interest. As shown in (a), three other groups are configured as parallel to the primary flat ($\angle\varphi = 90°$), $\angle\varphi = 30°$, and $\angle\varphi = 45°$ as examples, respectively. (b) Detail view of the three-sensor group G, where the five characteristic geometric parameters are illustrated. These geometric parameters need to be optimized based on sensitivity analysis. (c) Experimental system and circuit. The heaters (heater 1, heater 2) are energized individually by AC currents instead of synchronously during the test (i.e., heater 1 on & heater 2 off, or heater 1 off & heater 2 on), while the detector is energized by a constant DC throughout the measurement. This subfigure illustrates the circuit connection when the heater 1 (or heater 2) and the detector are working synchronously.

In this method, we merely adopt the second harmonic (2ω) voltage oscillation across the detector

($V_\text{D}^{2\omega}$) in each three-sensor group as the characteristic signal due to its good characteristics, hence the name "2ω method". Note that the commonly used third harmonic (3ω) voltage oscillations across the heaters are discarded in this method, which is explained in Sections 2.2 and 2.3. An AC current through the heater at frequency ω heats the sample at 2ω and produces the temperature oscillation at frequency 2ω, which makes the resistance of the heater and detector contain 2ω AC components. This resistance oscillation times the DC current results in a small 2ω voltage oscillation $V_\text{D}^{2\omega}$ across the detector.

Figure 1(b) shows the five characteristic geometric parameters of each three-sensor group that need to be engineered based on sensitivity analysis: the heater 1 width $w_\text{H1}$, the heater 1-detector distance $d_\text{H1D}$, the detector width $w_\text{D}$, the heater 2 width $w_\text{H2}$, and the heater 2-detector distance $d_\text{H2D}$. Sensitivity of the detector signal to the cross- and in-plane thermal conductivities ($\kappa_\text{cr}$, $\kappa_\text{in}$) can be adjusted directly by varying the heater width, the heater-detector distance, and the heating frequency $f_\text{H}$, which makes it feasible for the Intersection Method proposed in the next section. Specific guidelines for the layout design of each three-sensor group are detailed in Section 2.3 based on the sensitivity analysis in Supplementary Material (Section S1).

It is essential to select the length direction of each three-sensor group rationally. Owing to the large aspect ratio of each sensor (generally >15), the measured detector signal is insensitive to thermal conductivity parallel to the detector's length direction[28]. Hence, each three-sensor group is merely capable of deriving the cross-plane thermal conductivity, and the in-plane thermal conductivity perpendicular to the detector's length direction. In order to accurately derive the in-plane thermal conductivities along all orientations of interest, a corresponding number of directions of the three-sensor group should be arranged. As shown in Figure 1(a), an illustrative arrangement of four directions of the three-sensor group (∠$\varphi$ = 0, 30°, 45°, 90°) is designed to derive the in-plane thermal conductivities along the four corresponding directions. ∠$\varphi$ indicates the angle between the length direction of a three-sensor group and a given direction (e.g., the wafer secondary flat used in this figure). Note that if ones are interested in the overall in-plane thermal conductivity of a lateral isotropic material instead of the thermal conductivity along a specific orientation, the directions of the three-sensor group can be arbitrarily chosen.

Based on the three-sensor layout on the sample shown in Figure 1(a)-(b), the experimental system is built as Figure 1(c). First, the sample is assembled in a vacuum chamber to avoid the errors introduced by convection and radiation and to ensure accurate temperature control. Each three-sensor group is connected to the external circuit by a wire-bonding process, and the switching sequence of the two heaters is asynchronous. Specifically, the heater 1 is first energized individually by an AC current source (e.g., Keithley 6221) with the heater 2 power-off, while the detector is powered by a DC current source (e.g., Keithley 2450). And then, the heater 2 is energized individually by an AC current with the heater 1 power-off, while the detector continues to be powered by the DC current. The detector's 2ω signals are collected by a lock-in amplifier (e.g., SRS SR830), with the reference frequency and phase provided by the AC current source.

In addition, since the sample is covered with an insulation layer (e.g., amorphous $SiO_2$) or the sample itself is well-insulating, the 1ω voltage signals across the detector due to leakage from the heaters are negligible and are compatible with the lock-in amplifier's dynamic reserve. Hence, it is unnecessary to connect a variable resistor in series before the detector that subtracts the 1ω common-mode voltage signal as done in other existing 3ω-like methods[43, 45].

**2.2 Measurement Procedure**

The so-called "Intersection Method" is proposed here for measuring the cross- and in-plane thermal conductivities. The cross-plane thermal conductivity ($\kappa_{cr}$) and the in-plane thermal conductivity ($\kappa_{in}$) are derived by the intersection of the two $\kappa_{in}(\kappa_{cr})$ curves in ($\kappa_{cr}$, $\kappa_{in}$) coordinate, instead of being fitted by the multivariate fitting algorithms, which avoids the local optimum trap of multivariate fitting in principle. The two $\kappa_{in}(\kappa_{cr})$ curves are corresponding to the working sensor combinations (heater 1 & detector, or heater 2 & detector, respectively). The existence of the curve intersection is certain, provided the three-sensor layout is designed based on sensitivity analysis (as discussed in Section 2.3).

After the 2ω voltage signals across the DC detector ($V_D^{2\omega}$) are recorded by the lock-in amplifier, it is required to convert $V_D^{2\omega}$ into the 2ω temperature response signal of the DC detector ($\theta_D^{2\omega}$, the temperature oscillation amplitude), which is derived in the literature[38, 40] and our prior work[43, 46],

$$\theta_D^{2\omega} = \frac{\sqrt{2}\, V_D^{2\omega,\mathrm{rms}}}{I_D R_{D0}^{\mathrm{el}} \beta_D}. \tag{1}$$

The superscript "rms" denotes the root mean square, $I_D$ denotes the DC current across the detector, $R_{D0}^{el}$ denotes the detector's resistance at the reference temperature, and $\beta_D$ denotes the temperature coefficient of the detector's resistance (TCR).

Notice that the 3ω voltage signal across the AC heater ($V_H^{3\omega}$, corresponding to the temperature response of the AC heater $\theta_H^{2\omega} = \frac{2 V_H^{3\omega,\,rms}}{I_H^{rms} R_{H0}^{el} \beta_H}$) is discarded in this method, though it is commonly used in other 3ω-like methods[27, 28, 30]. Since the accurate a priori thermal properties of the insulation layer are not available, the high sensitivity of the AC heater's 3ω voltage signal to thermal properties of the superficial structures (insulation layer, insulation layer-sample interface, etc.) will introduce a large error into the final measured results. In contrast, the 2ω voltage signal across the DC detector at a certain distance from the heater is a promising solution, since several studies have unraveled that the sensitivity of the detector's 2ω voltage signal to thermal properties of the superficial structures can be significantly suppressed[38, 39, 43]. Therefore, the 2ω signal across the DC detector is merely adopted in this study. The sensitivity analysis corresponding to this issue is illustrated in Supplementary Material (Section S1).

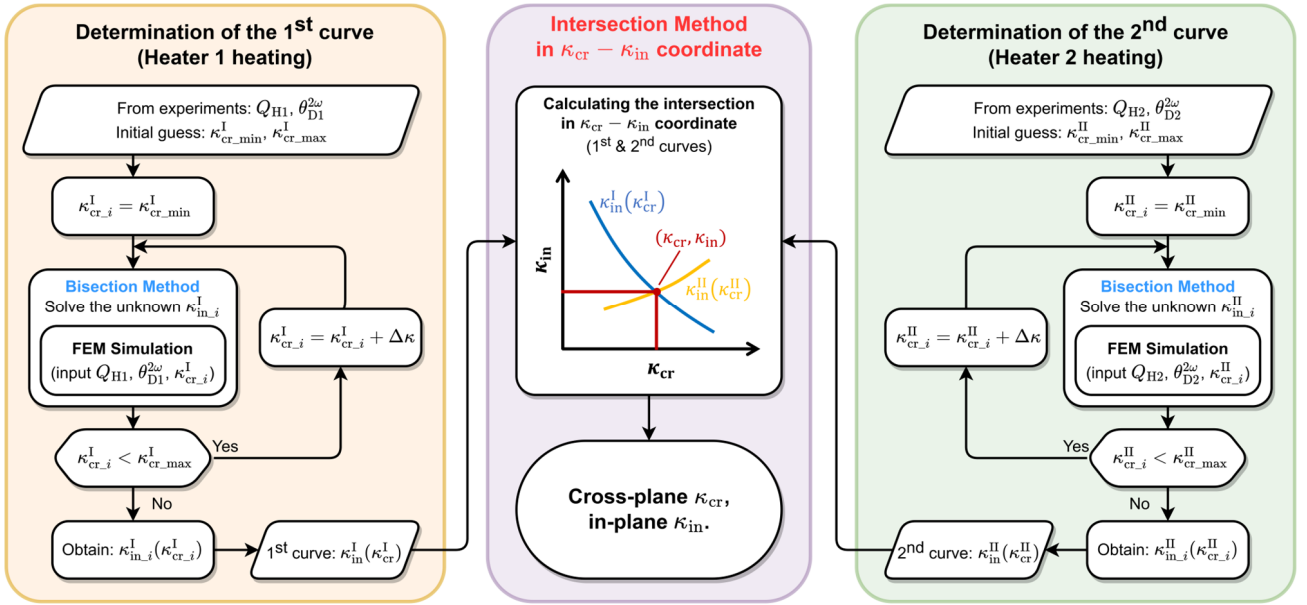

**Figure 2. The data processing procedure of the three-sensor 2ω method.** The core is the so-called "Intersection Method", which determines the cross- and in-plane thermal conductivities from the intersection of the two $\kappa_{in}(\kappa_{cr})$ curves. The existence of the curve intersection is certain, provided the three-sensor layout is designed based on sensitivity analysis. Note that the routine shown in this figure needs to perform on each three-

sensor group respectively to derive the anisotropic in-plane thermal conductivity along each direction of interest. The $\kappa_{in}$ and $\kappa_{cr}$ are mathematically equivalent, so one can arbitrarily adopt the $\kappa_{in}(\kappa_{cr})$ or $\kappa_{cr}(\kappa_{in})$ curve according to specific conditions.

Based on the sample structure and the experimental system shown in Figure 1, the heating power of the heater 1 ($Q_{H1}$), the heating power of the heater 2 ($Q_{H2}$), and the corresponding temperature responses of the detector ($\theta_{D1}^{2\omega}$, $\theta_{D2}^{2\omega}$) are measured, respectively. After that, thermal conductivities of the sample along each direction of interest can be derived by the following steps, i.e., the Intersection Method. Figure 2 illustrates steps (2) - (6) of the procedure.

(1) Construct a system in the finite element method (FEM) simulation that is consistent with the structure and boundary conditions of the sample.

(2) Set the heater 1's heating power ($Q_{H1}$) and the corresponding measured detector's temperature response ($\theta_{D1}^{2\omega}$) as the FEM simulation input.

(3) In FEM, continuously adjust the cross-plane thermal conductivity input ($\kappa_{cr}^{I}$), and then solve the nonlinear monadic equation $f(\kappa_{in}^{I}|\kappa_{cr} = \kappa_{cr}^{I}, Q_H = Q_{H1}) = \theta_{D1}^{2\omega}$ numerically (e.g., by Bisection Algorithm), to derive the unknown in-plane thermal conductivity ($\kappa_{in}^{I}$). Thus, the first curve $\kappa_{in}^{I}(\kappa_{cr}^{I})$ is obtained.

(4) Set the heater 2's heating power ($Q_{H2}$) and the corresponding measured detector's temperature response ($\theta_{D2}^{2\omega}$) as the FEM simulation input.

(5) In FEM, continuously adjust the cross-plane thermal conductivity input ($\kappa_{cr}^{II}$), and then solve the nonlinear monadic equation $f(\kappa_{in}^{II}|\kappa_{cr} = \kappa_{cr}^{II}, Q_H = Q_{H2}) = \theta_{D2}^{2\omega}$ numerically, to derive the unknown in-plane thermal conductivity ($\kappa_{in}^{II}$). Thus, the second curve $\kappa_{in}^{II}(\kappa_{cr}^{II})$ is obtained.

(6) Plot the two curves obtained in steps (3) and (5) in the ($\kappa_{cr}$, $\kappa_{in}$) coordinate, and their intersection indicates the measured cross- and in-plane thermal conductivities of the sample. Note that the in-plane thermal conductivity is normal to the length direction of the corresponding three-sensor group.

(7) For the three-sensor groups along different directions, repeat steps (2) to (6) to obtain the cross- and in-plane thermal conductivities corresponding to the length direction of each three-sensor group.

Indeed, it is easier and more efficient to implement the analytical solutions of harmonic heating rather than FEM simulations in this method. However, the available analytical solutions of harmonic heating[30, 47] are based on two main approximations: (1) 2D heat conduction, and (2) semi-infinite substrate, which are both unrealistic. In order to reproduce the same condition as experiments to ensure the accuracy, FEM simulations are performed here to overcome the two approximations of analytical solutions.

The FEM model are illustrated in in Supplementary Material (Section S4.4). In brief, the cross- and in-plane thermal conductivities are derived from the intersection of the two $\kappa_{in}(\kappa_{cr})$ curves in $(\kappa_{cr}, \kappa_{in})$ coordinate. Note that the $\kappa_{in}$ and $\kappa_{cr}$ are mathematically equivalent, so one can arbitrarily adopt the $\kappa_{in}(\kappa_{cr})$ or $\kappa_{cr}(\kappa_{in})$ curve according to specific conditions. Furthermore, if one wishes the intersection of two functions to be legible enough, the difference between the first-order derivatives of the two curves ($\frac{\partial \kappa_{cr}}{\partial \kappa_{in}}$, or equivalently the relative sensitivities $\frac{\partial \ln(\kappa_{cr})}{\partial \ln(\kappa_{in})}$) near the intersection point needs to be maximized. This lays a foundation for optimizing the three-sensor layout design in this method, which is detailed in the next section.

## 2.3 Three-sensor Layout Design

Guidelines for optimizing the three-sensor layout design and selecting the heating frequency are discussed in this section. In Section 2.2, we have stated that the solution to improving the accuracy of calculating intersection coordinates is to maximize the discrepancy of the first-order derivatives ($\frac{\partial \kappa_{cr}}{\partial \kappa_{in}}$, or the relative sensitivities $\frac{\partial \ln(\kappa_{cr})}{\partial \ln(\kappa_{in})}$) between the two curves $\kappa_{in}^{I}(\kappa_{cr}^{I})$ and $\kappa_{in}^{II}(\kappa_{cr}^{II})$ at the intersection, and the prerequisite for this strategy is optimizing the three-sensor layout design and selecting the heating frequencies.

The three-sensor layout design is determined by three characteristic geometric parameters: the heater width $w_H$, the heater-detector distance $d_{HD}$, and the detector width $w_D$, while the heating frequency $f_H$ is also an adjustable parameter. Hence, based on the detailed sensitivity analysis of sensors' signals and the resulting four feasible regions (Regions A, B, D, E shown in Figure S1 in Supplementary Material), guidelines for the three-sensor layout design and the selection of heating frequency are summarized as follows, and the corresponding schematic mind map is illustrated in Figure 3.

(1) To minimize $\frac{\partial \ln(\kappa_{cr})}{\partial \ln(\kappa_{in})}$: A wide heater 1 is arranged upon the sample surface, and a narrow detector 1 is located at a long distance from heater 1. Under this condition, the relative sensitivity ($\frac{\partial \ln(\kappa_{cr}^I)}{\partial \ln(\kappa_{in}^I)}$) of detector 1's 2ω signal approaches 0, with heater 1 individually heating the sample at a relatively high frequency.

(2) To maximize $\frac{\partial \ln(\kappa_{cr})}{\partial \ln(\kappa_{in})}$: A narrow heater 2 is arranged, and a narrow detector 2 is located at a medium distance from heater 2. Under this condition, the relative sensitivity ($\frac{\partial \ln(\kappa_{cr}^{II})}{\partial \ln(\kappa_{in}^{II})}$) of detector 2's 2ω signal approaches a large value, with heater 2 individually heating the sample at a low frequency.

(3) To restrain sensitivity to the effective thermal conductivity of insulation layer ($\kappa_{ins}$): The distance between heater 1(heater 2) and detector 1(detector 2) should not be too small.

(4) To simplify layout design: detector 1 and detector 2 can be merged into a single detector to simplify the sensor layout without affecting the measurement procedure and precision.

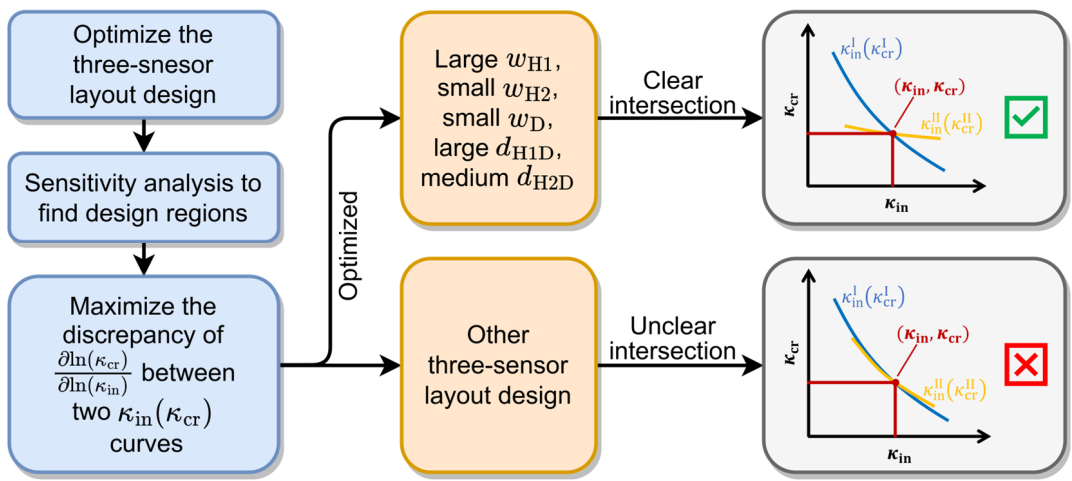

**Figure 3. Mind map of optimizing the three-sensor layout of the three-sensor 2ω method.** The key criterion is to maximize the discrepancy of $\frac{\partial \ln(\kappa_{cr})}{\partial \ln(\kappa_{in})}$ between the two curves (viz., $\kappa_{in}^I(\kappa_{cr}^I)$, $\kappa_{in}^{II}(\kappa_{cr}^{II})$) at the intersection, thereby making the intersection of two curves clear and legible. Here, $w_{H1}$, $w_{H2}$, $w_D$, $d_{H1D}$, $d_{H2D}$ are consistent with Figure 1(b), and the "optimized" parameter combination is a representative example within the feasible design space.

Note that the geometric parameter combination of three-sensor layout discussed here (i.e., large $w_{H1}$, small $w_{H2}$, small $w_D$, large $d_{H1D}$, medium $d_{H2D}$) is a representative example within the feasible design space, which is a handy design compatible with the stepper lithography and lift-off processes. Moreover, $\kappa_{ins}$ includes contributions from the sensor-insulation layer TBR, the thermal resistance of insulation layer, and the insulation layer-sample TBR[43].

Finally, the three-sensor layout consisting of a wide heater 1, a narrow heater 2, and a narrow detector, incorporating the appropriate distances and heating frequencies are engineered. As discussed before, the three-sensor design can eliminate the error propagation from uncertainties in the thermal resistance of superficial structures, and this is illustrated in Supplementary Material (Section S2, S5).

## 3. Results and Discussions

To verify the accuracy and reliability of the proposed three-sensor 2ω method, four typical mono-crystalline semiconductor substrates, i.e., (100) Si wafer, (0001) GaN wafer, (0001) AlN wafer, and (010) β-Ga$_2$O$_3$ substrate, are measured at 300 K. The four samples are prepared by different processes, among which the Si and β-Ga$_2$O$_3$ are fabricated by the Edge-defined Film-fed Growth (EFG) process[48], the GaN by the Hydride Vapor Phase Epitaxy (HVPE) process[49], and the AlN by the Physical Vapor Transport (PVT) process[50]. These wafers are purchased from corresponding companies.

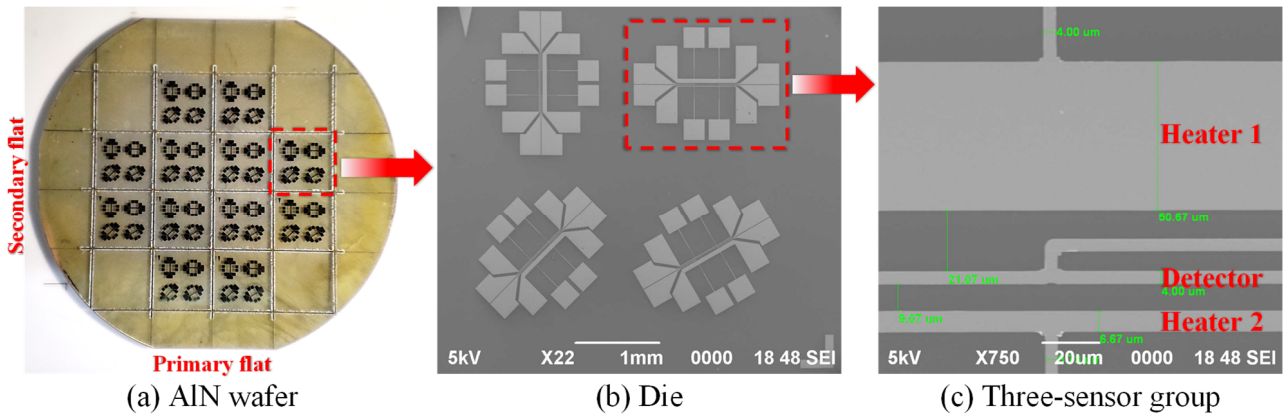

(a) AlN wafer      (b) Die      (c) Three-sensor group

**Figure 4. The practical three-sensor groups on the AlN wafer.** (a) The entire wafer with a primary flat (along a-axis, i.e., the [1-100] orientation) and a secondary flat (along m-axis). Twelve square dies with four three-sensor groups along different directions within them ($\angle\varphi = 0, 30°, 45°, 90°$) are fabricated upon the surface. (b) One of the dies on the wafer, and the triangle in the top left corner implies the direction normal to primary flat ($\angle\varphi = 0$). (c) Detail view of the three-sensor group parallel to the primary flat within a die.

To eliminate the leakage current and signal crosstalk between sensors, an amorphous $SiO_2$ insulation layer of ~40 nm is first deposited upon each sample by the Plasma Enhanced Chemical Vapor Deposition (PECVD) process[43]. Then the multiple three-sensor groups are fabricated upon the $SiO_2$ surface via lithography, magnetron sputtering, and lift-off processes successively[43, 46], as illustrated in Supplementary Material (Section S3.1), with the layouts and heating frequencies being optimized following the design guidelines depicted in Section 2.3. The material of each sensor is 90 nm Au/10 nm Cr (Cr for adhesion), and the actual patterns of the three-sensor groups are shown in Figure 4. For convenience, the material and deposition method of the insulation layer and of the metal sensors are all the same in four samples. Thus, the complete test samples are prepared.

The specific dimensions of three-sensor groups are characterized with the scanning electron microscope (SEM) and are detailed in Supplementary Material (Section S3.2). However, if one wants to measure a novel material with irregular shape and unavailable wafer, it is necessary to conduct an X-ray Diffraction (XRD) characterization to determine the feature crystal orientations of samples[51] before designing the sensor directions accordingly.

After the sample preparation and experimental circuit connection, the TCR ($\beta$) of all sensors need to be calibrated (Section S4.1 in Supplementary Material). After that, an AC current is first applied to the heater 1 (with frequency $f_{H1} = 2000$ Hz) in one of the three-sensor groups on the sample with the heater 2 power-off, and a constant DC current is applied to the detector. Next, an AC current is applied to the heater 2 ($f_{H2} = 300$ Hz) in the same three-sensor group with the heater 1 power-off, and the detector is continuously powered by the DC. The frequency of AC currents chosen for measuring all four samples are the same, whose universality is validated by the sensitivity analysis discussed in Supplementary Material (Section S1). The 2ω signals across the detector are recorded by a lock-in amplifier and converted into the temperature response signals ($\theta_{D1}^{2\omega}, \theta_{D2}^{2\omega}$) according to Eq. (1).

Then following the measurement procedure given in Section 2.2, a FEM simulation model is built as shown in Supplementary Material (Section S4.4), and the corresponding two curves $\kappa_{in}^{I}(\kappa_{cr}^{I})$, $\kappa_{in}^{II}(\kappa_{cr}^{II})$ can be calculated numerically (e.g., by the "fzero()" function in MATLAB®). Note that the $\kappa_{in}$ and $\kappa_{cr}$ are mathematically equivalent, so one can arbitrarily adopt the $\kappa_{in}(\kappa_{cr})$ or $\kappa_{cr}(\kappa_{in})$ curve for convenience.

In the FEM simulation, the thermal conductivity of each metal sensor is calculated from electrical conductivity by the Wiedemann-Franz Law as discussed in Supplementary Material (Section S4.2). The heat capacity and density of each material are referred to the SpringerMaterials database and the literature[19, 52], and the effective thermal conductivity of SiO₂ layer ($\kappa_{ins}$) is set to be 1.2 W/m K referring to our prior work[43], due to the same fab, growth process and SiO₂ structure. As explained before, $\kappa_{ins}$ includes the sensor-SiO₂ TBR, the thermal resistance of SiO₂, and the SiO₂-substrate TBR.

According to the Intersection Method, the cross-plane thermal conductivity and the in-plane thermal conductivity normal to the length direction of a three-sensor group are derived from the intersection coordinate of two $\kappa_{cr}(\kappa_{in})$ curves (as shown in Figure 5). By repeating the Intersection Method for all the three-sensor groups along different directions, comprehensive information about the cross- and in-plane thermal conductivities of the test sample can be obtained finally.

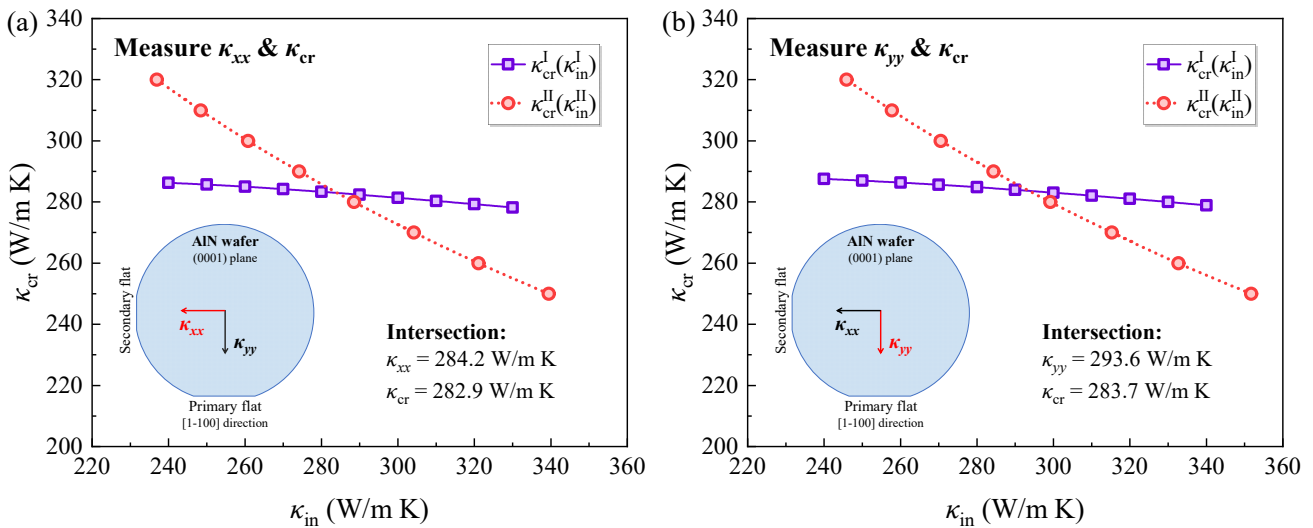

**Figure 5. Determine the cross- and in-plane thermal conductivities of the AlN wafer via the Intersection Method.** (a) Determine the cross-plane thermal conductivity $\kappa_{cr}$ and the in-plane thermal conductivity parallel to the primary flat $\kappa_{xx}$. (b) Determine $\kappa_{cr}$ and $\kappa_{yy}$. Average of the two $\kappa_{cr}$ measurements is chosen as the final $\kappa_{cr}$ result.

The measured thermal conductivities at room temperature are listed in Tables 2 and 3, and the error analysis is summarized in Supplementary Material (Section S5). For sample Si, GaN, and AlN, the thermal conductivities are all in good agreement with the literature[26, 53-66]. Specifically, the $\kappa_{xx}$, $\kappa_{yy}$, $\kappa_{zz}$ are along the orientations that parallel to the wafer primary flat (Si: [110], GaN or AlN: [1-100]), parallel to the wafer secondary flat (Si: [1-10], GaN or AlN: [11-20]), and normal to the wafer plane (Si: [001], GaN or AlN: [0001]), respectively. In addition, the thermal conductivities of these three samples are all approximately isotropic.

| Sample | Thermal conductivity (W/m K) | | | | |
|---|---|---|---|---|---|
| | $\kappa_{cr}$ | $\kappa_{xx}$ | $\kappa_{yy}$ | $\kappa_{ave}$ | $\kappa_{ref}$ |
| Si | 156.8 ± 4.4 | 146.5 ± 6.0 | 141.0 ± 5.8 | 148.1 ± 5.4 | 145 ± 7 [26] |
| GaN | 201.4 ± 5.6 | 203.7 ± 8.3 | 208.5 ± 8.5 | 204.5 ± 7.5 | 211 ± 5 [57] |
| AlN | 283.3 ± 7.2 | 284.2 ± 10.5 | 293.6 ± 10.8 | 287.0 ± 9.5 | 297 ± 7 [60] |

**Table 2.** Measured thermal conductivities of the three approximately isotropic materials (300 K). ($\kappa_{cr}$: cross-plane thermal conductivity, $\kappa_{xx}$: thermal conductivity parallel to the primary flat, $\kappa_{yy}$: thermal conductivity parallel to the secondary flat, $\kappa_{ave}$: effective bulk thermal conductivity)

Unlike the above three materials, the thermal conductivity of $\beta$-Ga$_2$O$_3$ is three-dimensional anisotropic, which stems from its monoclinic lattice[13, 67]. For the several characteristic crystal orientations (e.g., [010], [100]), the corresponding measurements are listed in Table 3. Based on the measured thermal conductivities along each main orientation, the components of the thermal conductivity tensor ($\kappa_{xx}$, $\kappa_{xy}$, $\kappa_{yy}$, and $\kappa_{zz}$) are derived according to the geometric relationships ($\kappa_{in}(\theta) = \kappa_{xx}\cos^2\theta + 2\kappa_{xy}\sin\theta\cos\theta + \kappa_{yy}\sin^2\theta$, $\kappa_{zz} = \kappa_{cr}$)[2, 19, 67]. Correspondingly, the entire thermal conductivity tensor of $\beta$-Ga$_2$O$_3$ can be determined as shown in Table 3. These results are consistent with the literature likewise[13, 19]. Our results are slightly higher than Jiang's[19], which may attribute to the differences in doping conditions (ours: undoped, Jiang's: Si-doped).

| Orientation | Thermal conductivity (W/m K) | | |
| --- | --- | --- | --- |
| | $\kappa$ | $\kappa$ tensor | $\kappa_{\text{ref}}$ |
| x-axis ([100]) | 10.5 ± 0.5 | | 9.5 ± 1.8 [19] |
| y-axis | 14.4 ± 0.8 | $\begin{bmatrix} 10.5 & -0.8 & 0 \\ -0.8 & 14.4 & 0 \\ 0 & 0 & 24.8 \end{bmatrix}$ | ≈ 13.3 ± 1.8 [19] |
| z-axis ([010]) | 24.8 ± 0.8 | | 22.5 ± 2.5 [19] |
| 45° to x, y-axis (near [102]) | 11.7 ± 0.6 | | ≈ 11.0 ± 1.8 [19] |

**Table 3.** Measured thermal conductivities of the sample $\beta$-Ga$_2$O$_3$ (300 K).

Figure 6 illustrates the comparison between the measured thermal conductivities in this study with other measurements and numerical simulations. In Figure 6(a), measured thermal conductivities of the four materials are laterally compared with the literature, and error bars are almost covered by data points. Moreover, $\kappa_{\text{in}}(\theta)$ of the $\beta$-Ga$_2$O$_3$ (010) plane are solely compared with the literature as shown in Figure 6(b), where our measured $\kappa_{\text{in}}(\theta)$ falls between those of Jiang et al.[19] and Guo et al.[13]. Meanwhile, our results are slightly lower than those of ShengBTE calculations based on the Gaussian approximation potential (GAP) calculated by Liu et al.[67], since the simulation neglects the lattice defects (e.g., isotope, unintentionally dope) distributing within the crystal. In summary, results obtained by the three-sensor 2ω method are consistent with the representative data available in the literature, which validates the accuracy and reliability of this method.

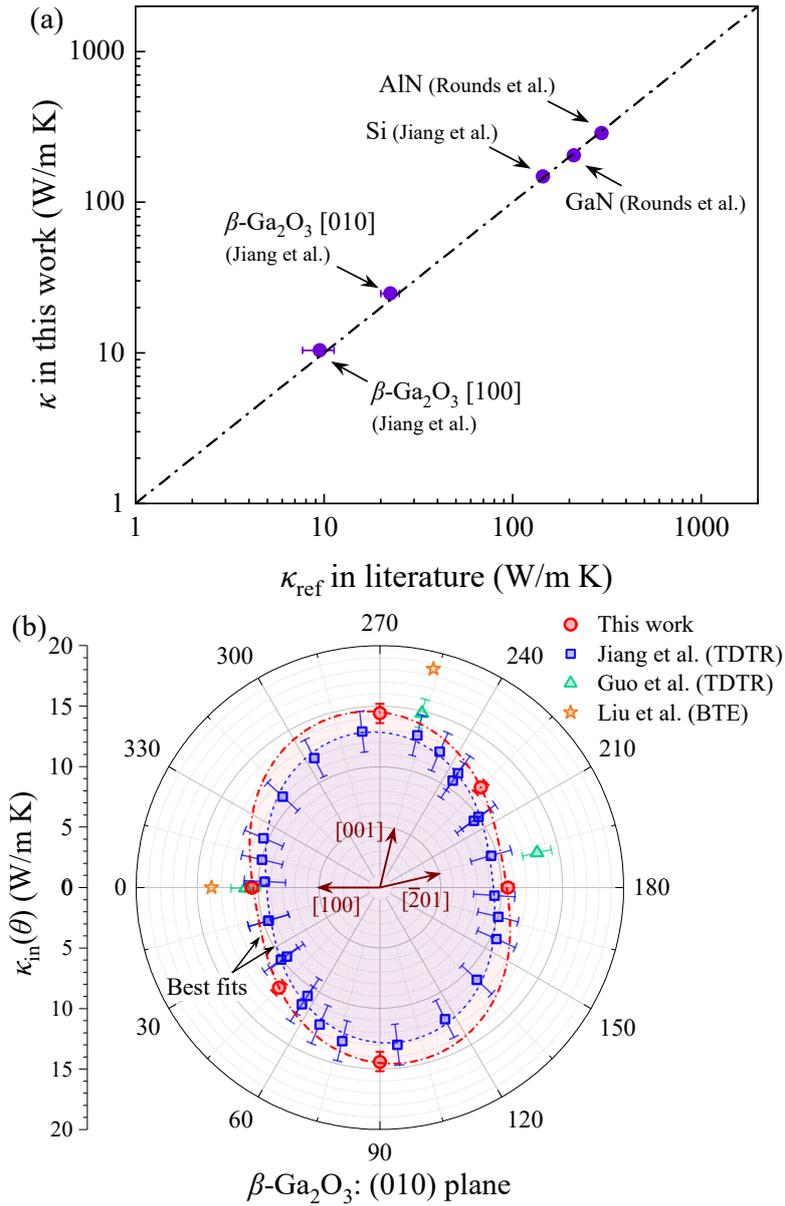

**Figure 6. Comparison between the measured thermal conductivities with the literature.** (a) A lateral comparison of the four materials' thermal conductivities with the literature. (b) A detailed comparison of in-plane thermal conductivity ($\kappa_{in}$) within the $\beta$-Ga$_2$O$_3$ (010) plane with the literature. The red dot dash ellipse shows the best fit of $\kappa_{in}(\theta)$ measured in this work, and is slightly higher than that of Jiang et al.[19] (blue dash ellipse) ascribing to different doping conditions. The orange marks show the ShengBTE calculations of mono-crystalline $\beta$-Ga$_2$O$_3$ based on the Gaussian approximation potential (GAP)[67].

## 4. Conclusions

In this work, we present the three-sensor 2ω method for measuring the thermal conductivity of solid materials, regardless of isotropy or anisotropy. To implement this method, multiple three-sensor groups are fabricated on the sample surface along different directions of interest. Each group contains three parallel metal sensors with optimized widths and distances according to design guidelines. The outer two sensors serve as AC heaters, while the middle sensor acts as a DC detector. The 2ω voltage signals across the detector are measured, and then data are processed using the proposed Intersection Method to obtain cross- and in-plane thermal conductivities along the directions of interest. Based on this method, four typical monocrystalline semiconductors, namely Si, GaN, AlN, and *β*-$Ga_2O_3$, are measured, and the results are consistent with the literature, verifying the accuracy and reliability of this method. The application of the detector's 2ω instead of the heater's 3ω signals eliminates the errors propagated from the uncertainties of thermal resistance in superficial structures (insulation layer, insulation layer-sample interface, etc.). In addition, this method replaces the commonly used multivariate fitting algorithms with the proposed Intersection Method, avoiding the local optimum trap of multivariate fitting. Therefore, the three-sensor 2ω method overcomes the main limitations of existing methods. This method will provide a comprehensive and universal solution for the thermal conductivity measurement of solid materials.

## Supplementary Material

See the Supplementary Material for the sensitivity analysis and the feasible regions for the three-sensor layout design (Section S1), the robustness analysis on the influence of insulation layer's thermal resistance (Section S2), the fabrication of the three-sensor layout (Section S3), the measurement details (Section S4), and the error analysis (Section S5).

## Acknowledgments

This work was financially supported by the National Natural Science Foundation of China (Grant No. 51825601, U20A20301).

[9]  L.-D. Zhao, S.-H. Lo, Y. Zhang, H. Sun, G. Tan, C. Uher, C. Wolverton, V.P. Dravid, M.G. Kanatzidis, Ultralow thermal conductivity and high thermoelectric figure of merit in SnSe crystals, Nature, 508(7496) (2014) 373-377.

[10] X. Yan, B. Poudel, Y. Ma, W.S. Liu, G. Joshi, H. Wang, Y. Lan, D. Wang, G. Chen, Z.F. Ren, Experimental Studies on Anisotropic Thermoelectric Properties and Structures of n-Type Bi2Te2.7Se0.3, Nano Letters, 10(9) (2010) 3373-3378.

[11] A.J. Schmidt, X. Chen, G. Chen, Pulse accumulation, radial heat conduction, and anisotropic thermal conductivity in pump-probe transient thermoreflectance, Review of Scientific Instruments, 79(11) (2008) 114902.

[12] X. Qian, P. Jiang, R. Yang, Anisotropic thermal conductivity of 4H and 6H silicon carbide measured using time-domain thermoreflectance, Materials Today Physics, 3 (2017) 70-75.

[13] Z. Guo, A. Verma, X. Wu, F. Sun, A. Hickman, T. Masui, A. Kuramata, M. Higashiwaki, D. Jena, T. Luo, Anisotropic thermal conductivity in single crystal β-gallium oxide, Applied Physics Letters, 106(11) (2015) 111909.

[14] P. Jiang, X. Qian, R. Yang, Time-domain thermoreflectance (TDTR) measurements of anisotropic thermal conductivity using a variable spot size approach, Review of Scientific Instruments, 88(7) (2017) 074901.

[15] A.J. Schmidt, R. Cheaito, M. Chiesa, A frequency-domain thermoreflectance method for the characterization of thermal properties, Review of Scientific Instruments, 80(9) (2009) 094901.

[16] E. Ziade, Wide bandwidth frequency-domain thermoreflectance: Volumetric heat capacity, anisotropic thermal conductivity, and thickness measurements, Review of Scientific Instruments, 91(12) (2020) 124901.

[17] L.A. Pérez, K. Xu, M.R. Wagner, B. Dörling, A. Perevedentsev, A.R. Goñi, M. Campoy-Quiles, M.I. Alonso, J.S. Reparaz, Anisotropic thermoreflectance thermometry: A contactless frequency-domain thermoreflectance approach to study anisotropic thermal transport, Review of Scientific Instruments, 93(3) (2022) 034902.
20 / 25